\documentclass[10pt,letterpaper]{article}
\usepackage{ol21}
\usepackage{graphicx}
\usepackage[T1]{fontenc}
\usepackage{subfigure}
\usepackage{amsmath}
\usepackage{amsfonts}
\usepackage{stmaryrd}
\usepackage{psfrag}
\usepackage{tabularx}\newcolumntype{Y}{>{\centering\arraybackslash}X}

\begin{document}


\title{Remote optical sensing on the nanometer scale with a bowtie aperture nano-antenna on a SNOM fiber tip} 



\author{Elie M. Atie$^{1,2}$,Zhihua Xie$^{1}$, Ali El Eter$^{1}$, Roland Salut$^{1}$,  Dusan Nedeljkovic$^{3}$,Tony Tannous$^{2}$ Fadi I. Baida$^{1}$ and Thierry Grosjean$^{1}$*}

\address{$^{1}$ Institut FEMTO-ST, UMR CNRS 6174, Universit\'e  de Franche-Comt\'e, D\'epartement d'Optique P.M. Duffieux,  15B avenue des Montboucons, 25030 Besan\c{c}on cedex, France}

\address{$^{2}$ Department of Physics, University of Balamand, Lebanon}

\address{$^{3}$ Lovalite s.a.s., 7 rue Xavier Marmier, 25000 Besan\c{c}on, France.}

\email{* thierry.grosjean@univ-fcomte.fr}


\begin{abstract}
Plasmonic nano-antennas have proven the outstanding ability of sensing chemical and physical processes  down to the nanometer scale. Sensing is usually achieved within the highly confined optical fields generated resonantly by the nano-antennas, i.e. in contact to the nanostructures.
In this paper, we demonstrate the sensing capability of nano-antennas to their larger scale environment, well beyond their plasmonic confinement volume, leading to the concept of ``remote'' (non contact) sensing on the nanometer scale. On the basis of a bowtie-aperture nano-antenna (BNA) integrated at the apex of a SNOM fiber tip, we introduce an ultra-compact, moveable and background-free optical nanosensor for the remote sensing of a silicon surface (up to distance of $300~nm$). Sensitivity of the BNA to its large scale environment is high enough to expect the monitoring and control of the spacing between the nano-antenna and a silicon surface with sub-nanometer accuracy. This work paves the way towards a new class of nanopositionning technique, based on nano-antenna resonance monitoring, that are alternative to nanomechanical and optical interference-based devices.
\end{abstract}


\maketitle 

\section{Introduction}

Plasmonic nano-antennas (NA) have emerged as a powerful tool for sensing and monitoring variations of their local environment such as the binding of molecules at the nano-antenna surface\cite{lal:natphot07,anker:natmat08,stewart:chemrev08}, hydrogen gas absorption\cite{liu:natmat11}, chemical reactions \cite{larsson:science09,tittl:nl13}, phase-changing processes\cite{lei:ol10}. Such sensing properties are due to the frequency shift capability of localized surface plasmon resonances under tiny changes in the effective dielectric constant of the surrounding medium. In all the above-cited examples, nano-antennas have monitored chemical and physical processes in contact to the nano-antennas, by taking advantage of the highly localized plasmon fields that are resonantly generated within of the nanostructures. The sensing ability of nano-antennas to changes happening at a distance from the nanostructures has been hardly discussed. A hybrid structure combining a plasmonic nano-antenna and a dielectric Fabry Perot cavity have been demonstrated to remotely sense with nanometer spatial accuracy important film parameters in thin-film nanometrology \cite{schmidt:natcomm12} but, to our knowledge, no investigation of remote sensing with single plasmonic nano-antennas have been reported so far. Moreover, applications of NA-based refractive index sensing are restricted to the extraction of weak nanoscale plasmonic information from a strong background signal. Direct background-free optical detection of plasmonic nanosensors is highly desirable because it would avoid the use of complex detection mechanisms such as absorption/extinction or dark-field scattering spectroscopies \cite{liao:fm06,schmidt:natcomm12,liu:natmat11}..

This letter addresses the concept of fiber-integrated bowtie aperture nano-antenna (BNA)\cite{mivelle:ox10,vo:ox12,mivelle:nl12,mivelle:ox14,eleter:ox14,berthelot:natnanotech14} as a metallic nanoresonator for remote sensing applications. This aperture-type nano-antenna \cite{jin:apl05,wang:nl06} opened at the end of an opaque metal coated SNOM fiber tip offers the unique ability of a direct in-fiber and background-free detection of the far-field optical signal emitted by the BNA under resonance. This moveable nano-antenna is demonstrated in the remote (i.e., non-contact) sensing of a silicon surface with direct in-fiber detection, bringing innovative solution to nanopositioning. Given the extremely high sensitivity of BNAs to changes of their environment, the tip-to-surface distance can be monitored and controlled over a range at least one order of magnitude larger than that of mechanical force detection devices used in shear-force and atomic force microscopy (AFM), with sub-nanometer positionning accuracy and subwavelength lateral sensing area.

\section{BNA on fiber SNOM tip}
\begin{figure}[htbp]
\begin{center}
\includegraphics [width=0.5\columnwidth]{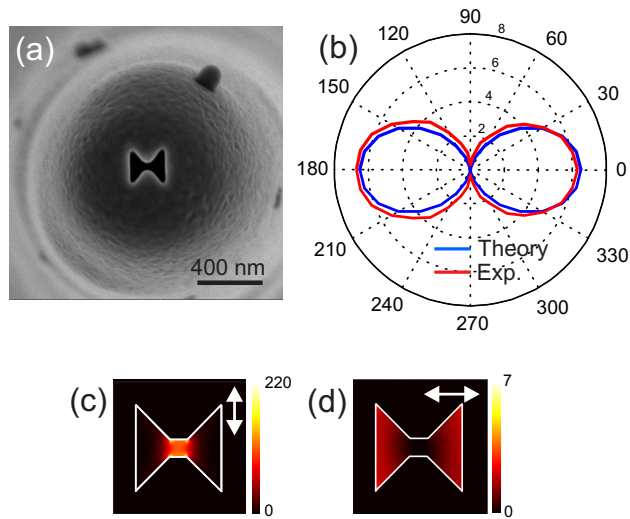}
\caption{(a) SEM micrograph of a BNA on a SNOM fiber tip. (b) Experimental and ideal polarization diagrams of the BNA-on-tip. (c,d) Simulation of the enhancement of optical electric intensity plotted in the transverse plane taken at half the thickness of the metal layer right at the feed gap of the BNA ($\lambda$=1.55 $\mu$m), for two perpendicular polarization directions of the incoming wave (see arrows). This enhancement is the ratio between the field intensity diffracted by the BNA placed at 2 nm beyond a Si substrate and the intensity calculated without the presence of the tip.\label{fig:BNA}}
\end{center}
\end{figure}

In order to produce the BNA-on-tip, polymer tips are first grown by photopolymerization at the cleaved end facet of a monomode ($1550~nm$ wavelength) glass fiber \cite{bachelot:ao01}. Next, the probes are metal-coated with $100~nm$ thick aluminum layer. Finally, the BNA is fabricated at the apex of the tip by FIB milling (dual beam FEI Helios 600i with a Raith Elphy Multibeam attachment). Figure \ref{fig:BNA}(a) displays scanning electron micrographs of a resulting fiber device.

Two BNAs of different sizes are considered in this study. The first one (BNA1) is $210~nm$ wide with a square gap about $40~nm$ large and $45^{\circ}$ flare angles (see Fig.\ref{fig:BNA}(a)). With this geometry, the BNA's optical resonance is centered to a wavelength of $\lambda=1500~nm$  when it is located at a distance of $5~nm$ away from a silicon substrate $(n=3.4)$. The second one (BNA2) has a smaller gap of about $20~nm$ and is $250~nm$ large so that its resonance is redshifted with respected to the first BNA, centered to $\lambda=1850~nm$ at $5~nm$ far from the Si surface. The design process is performed using three-dimensional Finite Difference Time Domain method (3D FDTD) \cite{taflove:book}. Details on the polarization-dependent resonance process of the BNA can be found in Refs. \cite{jin:apl05,mivelle:ox10,mivelle:ox14}. We just note that the resonant mode of the BNA is strongly bound to its nanometer-scale feed gap which strongly enhances the optical electric field by optical capacitive effect (see Fig. \ref{fig:BNA}(c) for the first BNA). The extent of the resulting light confinement beyond the tip apex does not exceed $40~nm$ ($1/e^2$ calculation of the longitudinal intensity profile along the tip axis). This tiny optical "hot spot" vanishes  when the input polarization is turned by $90^{\circ}$, due to resonance cancelation (cf. Fig. \ref{fig:BNA}(d) for the first BNA). Figure  \ref{fig:BNA}(b) displays the polarization diagram of the first BNA on tip used in collection mode (red curve). A weakly focused linearly polarized beam is directly projected onto the nano-antenna and the signal collected by the tip (which is due exclusively to the BNA resonance) is measured at the fiber end as a function of the incident polarization direction. We see that this diagram shows a two-lobes structure which fits the polarization diagram of an ideal oriented dipole detector described by a cosine function (blue curve). This proves that the BNA can be described optically by an electric dipole moment. The first BNA shows a polarization ratio of about 1:20. The polarization ratio of the second BNA is measured to a value of 1:8.

\section{Experimental set-up}

These two nano-antennas are approached to a Si surface and studied optically with the experimental set-up shown in Fig. \ref{fig:scheme}(a). The BNAs on fiber tip are mounted onto a tuning fork and placed into a SNOM head from NT-MDT company (NTEGRA System). A shear-force detection based feedback loop is used to position the tip in close proximity of the surface of a Si wafer. The Si sample is mounted onto a calibrated piezo stage (PI company) to be accurately moved away from the tip along the longitudinal direction (perpendicular to the sample surface). The BNA on tip is used in collection mode. Laser light at $1550~nm$ ($4~mW$) is slightly focused with a lens onto the tip apex through the Si wafer. The fiber output is coupled to an InGaAS photodiode (Thorlabs) and a synchronous detection scheme is used to enhance signal-to-noise ratio. The output voltage is directly monitored with a numerical oscilloscope connected to a computer. The BNA is resonantly excited by aligning the polarization direction of the incident optical waves along the BNA's polarization axis, ie. the symmetry axis of the BNA that passes through each metal triangle's tip.

\section{Results and discussions}

\begin{figure}[htbp]
\begin{center}
\includegraphics [width=0.99\textwidth]{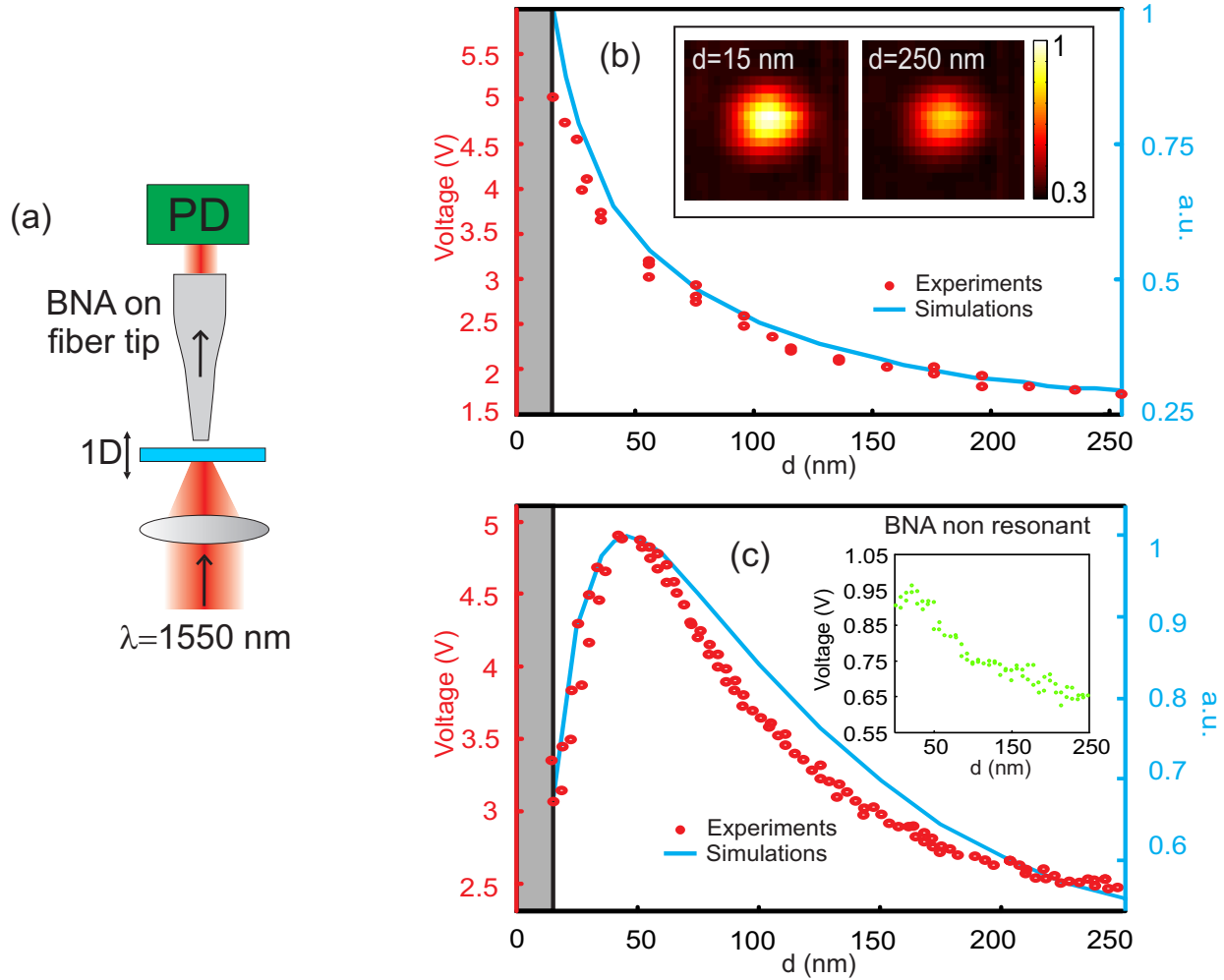}
\caption{(a) Scheme of the experimental set up. (b,c) Approach curves towards a Si surface obtained at $\lambda=1550~nm$ with (b) the first and (c) the second BNA-on-tip: experimental results (circles) and numerical predictions (solid line). The BNA-on-tips are used in collection mode. Inset of (b): far-field images of the first BNA used in emission mode (light is injected into the fiber and the BNA is imaged with an objective coupled to an infrared camera). Inset of (c): experimental approach curve with a non-resonant excitation of the second BNA (the incident polarization is perpendicular to the BNA's polarization axis).}\label{fig:scheme}
\end{center}
\end{figure}

Figure \ref{fig:scheme} (b) and (c) show the optical signal collected by the two BNA-on-tips considered in this study, as a function of the spacing ($d$) between the nanostructures and the surface.  To this end, the tip is set in close proximity of the sample with shear force distance control facility. Then, the sample is moved away from the surface by steps of a few nanometers with the piezo stage and at each step the signal is recorded. The minimum spacing, for which the feedback loop is triggered, is known to be about $15~nm$$\pm$ $5~nm$. To preserve the tip, we did not investigate this zone, this is why the last $15~nm$ before contact appear as a shaded area in the figures. The experimental approach curves (circles) are compared with a numerical analysis of the problem (solid line). In the simulation, the BNA is excited with an incoming plane wave in the Si substrate at normal incidence that is polarized along the BNA's polarization axis. Data are obtained with Poynting vector flow calculation through the output cross-section of the 2-micron-long tip portion considered in the simulation.

\begin{figure}[htbp]
\begin{center}
\includegraphics [width=0.99\textwidth]{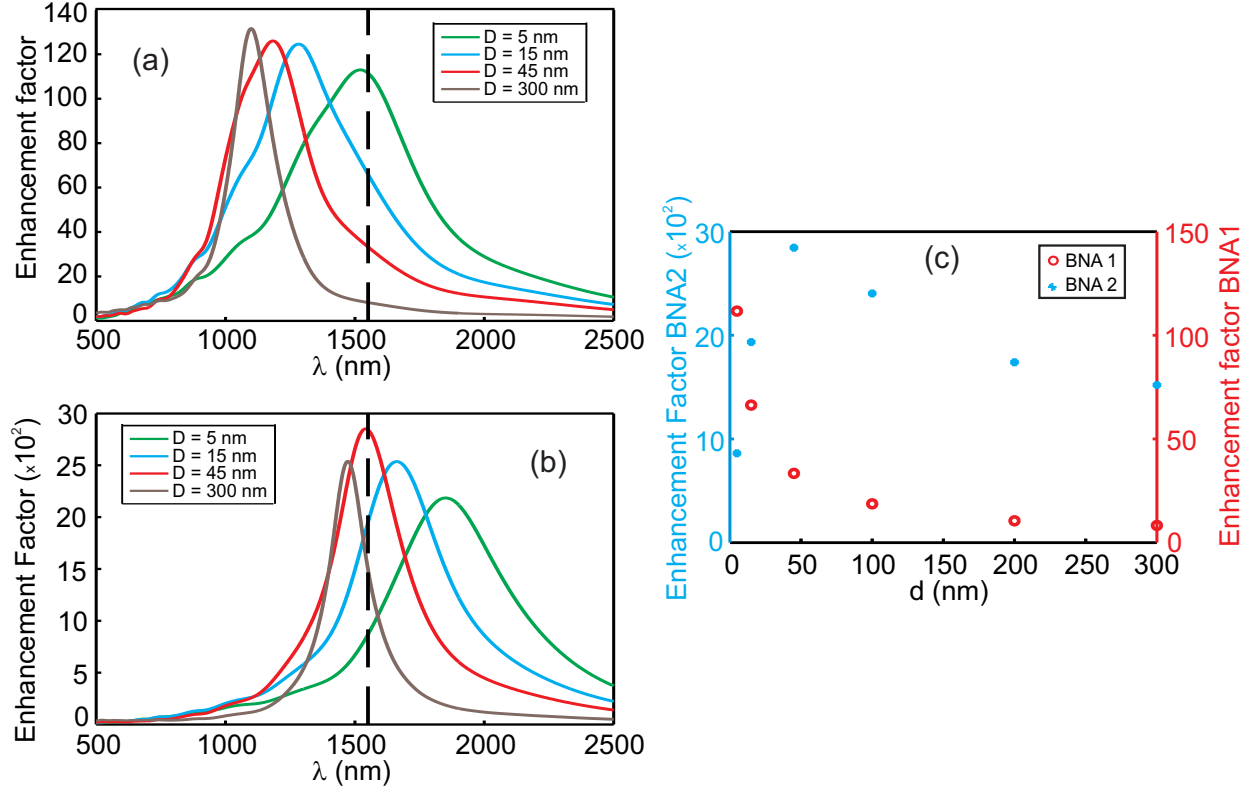}
\caption{(a,b) Simulation of the normalized resonance spectra of the first and second BNAs close to a Si substrate, respectively, for various BNA-to-surface spacings. (c) Plots of the intensity enhancement at the center of the first (solid line) and second (dashed line) BNAs, as a function of the tip-to-surface spacing.}\label{fig:spectre}
\end{center}
\end{figure}

We see that the approach curves obtained with the two BNA configurations show important discrepancies. With the first BNA (Fig. \ref{fig:scheme}(b)), the collected signal increases continuously when the spacing $d$ decreases, whereas for the second BNA (Fig. \ref{fig:scheme}(c)), the collected signal is increased for decreasing spacings down to 45 nm and is then strongly decreased for smaller distances. The origin of this discrepancy is evidenced in the spectral analysis of the two BNAs shown in Fig. \ref{fig:spectre}. Figures \ref{fig:spectre}(a) and (b) show the theoretical resonance spectra of the BNAs at proximity of a Si substrate, for various values of BNA-to-surface spacings $d$ smaller than $300~nm$. The BNAs are excited with a pulsed incoming plane wave in the Si substrate at normal incidence. The spectra are obtained by Fourier transforming the electric field recorded during the simulation at a single grid cell, located at the center of the BNAs.  They are normalized by the spectra calculated at the same point without the tip to provide information about intensity enhancement. The experimental wavelength $\lambda$=1550 nm is emphasized with dashed lines into the figures. We see in Figs. \ref{fig:spectre}(a) and (b) that the presence of a silicon surface close to the BNAs leads to a dramatic redshift of the nano-antenna resonance that exceeds $400~nm$.  For a spacing varying from $300~nm$ down to $5~nm$, the resonance wavelength shifts from $1100~nm$ to $1500~nm$ with the first BNA (Fig. \ref{fig:spectre}(a)) and from $1400~nm$ to $1850~nm$ with the second BNA (Fig. \ref{fig:spectre}(b)). Therefore, with the first BNA, the resonance peak stays centered to wavelengths smaller than $1550~nm$ over the experimental values of $d$ comprised between $300~nm$ and $15~nm$. This results, at a wavelength of $1550~nm$, in an intensity enhancement at the BNA's gap that continuously increases while the tip approaches the Si surface. Such a phenomenon is clearly shown in Fig. \ref{fig:spectre}(c) which displays the value of the spectra of the two BNAs at $\lambda=1550~nm$, for various value of $d$. Given the electric dipolar properties of the BNA, optical emission into the tip should be continuously increased during the tip approach, which is consistent with Fig. \ref{fig:scheme}(b). The inset of Fig. \ref{fig:scheme}(b)  show far-field images of the first BNA-on-tip used in emission mode, for two tip-to-surface spacings of $15~nm$ (right inset) and $250~nm$ (left inset).  To this end, laser light $(50~\mu$W$)$ is injected into the fiber so that the BNA is resonantly excited and the magnified far-field images of the BNA through the Si wafer are realized with a long working distance objective ($\times$60, 0.7) coupled to an infrared camera. We see that the light spot intensity right at the BNA is increased close to the surface, which means that the overall BNA emission is modified by the sample, not only the BNA emission into the tip. This property avoids changes in the collected signal due to modifications of the dipolar emission pattern of the BNA close to a high index dielectric substrate \cite{lukosz:josa79}. Note that this study stays qualitative since IR cameras undergo noticeable non-linearity in their optical response. From Fig. \ref{fig:spectre}(b), we see that the resonance peak of the second BNA is centered to $\lambda=1550~nm$ when it is positioned $45~nm$ away from the Si surface. As a result, the electric field enhanced by the nano-antenna at $\lambda=1550~nm$ during tip approach towards the sample surface is maximum at $d=45~nm$, as shown in Fig. \ref{fig:spectre}(c), which is fully consistent with experimental results given the electric dipole emission properties of the BNA. When the BNA is excited off-resonance, by rotating the incident polarization by $90^{\circ}$, the approach curve does not show maximum anymore and the collected signal is 6 times weaker (see inset of Fig. \ref{fig:scheme}(c)). The approached curve is described by an exponential like function typical to that Near field microscopy. Therefore, variations of the collected signal during tip approach are mainly governed by the resonance redshift of the BNA, rather than other optical processes that could be in play such as tip-to-surface interferences. These variations, strongly dependent on BNA geometry, are thus the signature that the BNA is capable of sensing the presence of a flat sample even when it is located at distances from the sample surface well beyond its resonant optical confinement ($1/e$ field amplitude).

Nano-positioning techniques used in scanning probe microscopies are generally based on processing low voltage variations that results from the detection of mechanical or optical signals. In our case, the detection voltage range has been set to a few volts together with low noise with the synchronous detection scheme ($10~ms$ integration time). In that case, the tangential slope of the experimental curves of Fig. \ref{fig:scheme}(b) is measured to values about $-119~mV.nm^{-1}$, $-32~mV.nm^{-1}$, $-15~mV.nm^{-1}$ and $-5~mV.nm^{-1}$ at tip-to-sample spacings of $15~nm$, $50~nm$, $100~nm$ and $200~nm$, respectively. The second experimental curves of Fig. \ref{fig:scheme}(c) show local tangential slopes about $233~mV.nm^{-1}$, $46~mV.nm^{-1}$, $-19~mV.nm^{-1}$ and $-4~mV.nm^{-1}$ at tip-to-surface spacings of $15~nm$, $30~nm$, $100~nm$ and $200~nm$, respectively. Since the millivolt range is beyond the noise level of the current highly sensitive optical detection devices, one can expect with BNA-on-tips distance control schemes of sub-nanometer accuracy along the longitudinal direction perpendicular to the sample surface. The on-tip aperture type nature of the BNA leads to surface sensing in subwavelength areas, in compact, fiber and plug-and-play optical architectures that require modest power consumption.  Surface sensing is usually achieved on the nanometer scale by using mechanical probes aimed at detecting locally surface forces (shear force and atomic force microscopy). If transverse  spatial resolution can be of the order of the angstrom, the spatial sensing depth is limited to a few nanometers. In our case, one can expect distance regulation over distances at least two orders of magnitude higher and with subwavelength sensing lateral area.

\section{Conclusion}

In our paper we have demonstrated the remote sensing capability of using a BNA engineered at the end of a SNOM fiber probe. The BNA has been shown to sense the presence of a Si surface up to distance of $300~nm$. We thus prove that NA's have the ability of detecting variation in their large scale environment. Such sensing is due to the high sensitivity of the BNA to a small variation in the effective dielectric constant of the medium surrounding it. In addition, our experimental result have presented a high agreement with the theoretical one, and showed good voltage variation relative to the variation of the distance. Consequently, a single BNA engraved at the apex of a SNOM tip may represent an alternative solution for nano-positioning and distance control instead of nano-mechanical and optical interference-based techniques used nowadays 

\section{Acknowledgments}
This work is funded by the Agence Nationale de la Recherche (ANR) under contract  NANOEC (ANR-07-NANO-036) and Baltrap (ANR-10-NANO-002). It is supported by the  ``P\^ole de comp\'etitivit\'e Microtechnique'', the Labex ACTION, and the french network of technology platforms ``Renatech''. Tip metal coating and FIB fabrication are realized at MIMENTO technology platform.


\end{document}